\documentclass[aps,twocolumn,prl, superscriptaddress]{revtex4-2}

\usepackage[english]{babel}
\usepackage{amsmath, amssymb, mathtools, physics}
\usepackage{braket}
\usepackage{ifthen}
\usepackage{graphicx,float}
\usepackage{multirow}
\usepackage[RGB,dvipsnames]{xcolor}
\definecolor{dodgerblue}{rgb}{0.12, 0.56, 1.0}
\definecolor{darkcyan32144140}{RGB}{32,144,140}
\definecolor{darkslateblue5294141}{RGB}{52,94,141}      
\definecolor{darkslateblue6467135}{RGB}{64,67,135}
\definecolor{darkslateblue7235116}{RGB}{72,35,116}      
\definecolor{darkslategray38}{RGB}{38,38,38}
\definecolor{greenyellow18922238}{RGB}{189,222,38}      
\definecolor{indigo68184}{RGB}{68,1,84}                 
\definecolor{lavender234234242}{RGB}{234,234,242}
\definecolor{mediumseagreen34167132}{RGB}{34,167,132}   
\definecolor{mediumseagreen68190112}{RGB}{68,190,112}
\definecolor{teal41120142}{RGB}{41,120,142}
\definecolor{yellowgreen12120981}{RGB}{121,209,81}
\colorlet{bluedarkslateblue5294141}{blue!50!darkslateblue5294141}
\newcommand{\re}[1]{{\leavevmode#1}}
\newcommand{\rev}[1]{{\leavevmode#1}}

\newcommand{\eprM}{\sigma_\text{M}}
\newcommand{\eprSM}{\sigma_\text{WTD}}
\newcommand{\eprTUR}{\sigma_\text{TUR}}
\newcommand{\eprMTUR}{\sigma_\text{MTUR}}

\newcommand{\superscriptProcessing}{\text{pm}} 

\newcommand{\wtdpsitheo}[3]{\psi^\text{th}_{#1\to #2}(#3)}
\newcommand{\wtdpsiexp}[3]{\psi^\text{ex}_{#1\to #2}(#3)}
\newcommand{\wtdpsips}[3]{\psi^\superscriptProcessing_{#1\to #2}(#3)}
\newcommand{\wtdpsiss}[3]{\psi^\text{ss}_{#1\to #2}(#3)}
\newcommand{\nrateexp}[1]{\nu^\text{ex}_{#1}}
\newcommand{\nrateps}[1]{\nu^\superscriptProcessing_{#1}}

\newcommand{\currentps}[1]{j^\superscriptProcessing_{#1}}

\newcommand{\alletas}{\boldsymbol{\eta}}
\newcommand{\etastar}[1]{\eta_{#1}^\ast}
\DeclareMathOperator{\Var}{Var}

\newcommand{\cumu}{\mathcal{C}}

\newcommand{\paragraphLetter}[1]{\textit{#1}.---\!\!}
\usepackage{hyperref}
\usepackage[nameinlink]{cleveref}
\Crefname{equation}{Eq.}{Eqs.}
\Crefname{figure}{Fig.}{Figs.}
\usepackage{enumitem}
\usepackage{tikz}
\usetikzlibrary{external,cd}
\tikzsetexternalprefix{tikz/}
\usepackage{pgfplots}
\usepgfplotslibrary{colormaps}
\renewcommand{\vec}{\boldsymbol}
\usetikzlibrary{automata, arrows.meta, positioning, calc,shapes}

\begin{document}
%
\title{Compensating random transition-detection blackouts in Markov networks}
\author{Alexander M. Maier}
\author{Benjamin H\"asler}
\author{Udo Seifert}

\affiliation{%
 II. Institut für Theoretische Physik, Universität Stuttgart, 70550 Stuttgart, Germany
}%

\begin{abstract}
In Markov networks, measurement blackouts with unknown frequency compromise observations such that thermodynamic quantities can no longer be inferred reliably. In particular, the observed currents neither discern equilibrium from non-equilibrium nor can they be used in extant estimators of entropy production. Our strategy to eliminate these effects is based on formally attributing the blackouts to a second channel connecting  states. The unknown frequency of blackouts and the true underlying transition rates can be determined from the short-time limit of observed waiting-time distributions. A post-modification of observed trajectory data yields a virtual effective dynamics from which the lower bound on entropy production based on thermodynamic uncertainty relations can be recovered fully. Moreover, the post-processed data can be used in waiting-time based estimators. Crucially, our strategy does \re{not require the blackouts to occur homogeneously or symmetrically under time-reversal}.
\end{abstract}
\maketitle

\paragraphLetter{Introduction}
Stochastic thermodynamics provides a comprehensive framework for the description and analysis of classical non-equilibrium systems on the nano- and micro-scale \cite{seki10,jarz11,peli21,shir23,seif25}. Its main assumption of an underlying stochastic Markovian dynamics is well verified experimentally for a huge class of systems ranging from electronic systems \cite{peko15} to colloids \cite{cili17} and simple biomolecules \cite{alem15,kolo15,gnes18}. In more complex systems, however, not all of the relevant degrees of freedom participating in this underlying Markovian dynamics are experimentally accessible. At this point, the strategy of thermodynamic inference comes on stage aiming to extract information about thermodynamic quantities -- with entropy production arguably being the most prominent one --  from  observations that explicitly or implicitly involve coarse-graining over the underlying degrees of freedom \cite{seif19,dieb25}.

Model-free lower bounds on entropy production can be based on the observation of transitions between lumped states \cite{espo12,bo14,skin21}, on evaluating the ratio of observed  probabilities of forward and time-reversed coarse-grained trajectories \cite{kawa07,gome08a,rold10}, on the asymmetry of correlation functions \cite{dech23,dieb25a,vu23}, on measuring the statistics of certain rare events \cite{rold15,neri17,pigo17,neri23}, on determining the displacement and force variances through the variance sum rule \cite{dite24}, on the measurement of currents and their variances through the thermodynamic uncertainty relation (TUR) \cite{bara15,ging16,hyeo17,hwan18,song21,magg23} and its variants and refinements \cite{proe17,dech18,busi19,liu19,horo20,koyu20,dech20b,dech20a,mani20,otsu20,vanv20a,dech21,rold21,otsu22,dieb23,vu23a,piet24,mani24,asly25,indo25} or speed limits \cite{yosh21,ito17,shir18,vo20,fala20a,ito20}, and on the waiting-time distributions (WTDs) of observed transitions \cite{mart19,skin21a,ehri21,vdm22,haru22,vdm22b,blom24a,maie24,haru24b,kapu22,erte24,degu24,maie25}.

In essentially all of these techniques, a necessary requirement is that the observations are complete and error-free\re{, the latter being a requirement recent works attempt to overcome \cite{ferr25,vdm25,bao25}}. Complete here means that once it is defined which (lumped) states, sequence of states, and transitions are ``observable'' in a given scheme, the observation will record completely whenever the system is in these states and undergoes the respective transitions. In other words, the observer does not encounter blackouts where the recording misses a state or a transition that is, in principle, observable.

Such blackouts are potentially devastating since they may mislead an observer even in the qualitative distinction between equilibrium and non-equilibrium. For a simple example, consider a three-state system where we can observe transitions through one of its links in the forward (+) and the backward (--) direction. If this observation is complete, then counting the frequency of + and -- transitions suffices to distinguish equilibrium (same frequency) from non-equilibrium (a net current in one direction). If, however, we see only  fractions $\eta_+$ and $\eta_-\not = \eta_+$ of the forward and the backward transitions, respectively, then we will falsely assign non-equilibrium even to an equilibrium system. Likewise, for a non-equilibrium system, we will measure a mean current and its variance that, if used naively in the TUR, may even overestimate the entropy production.

The purpose of this Letter is to introduce an approach that allows us to compensate for such unknown time-asymmetric random blackouts in the observation of transitions between the states of a Markov network. We will combine an analysis of the short-time limit of waiting-time distributions between the observed transitions with post-modification of observed trajectories. This strategy allows us first to determine for each link the unknown fractions of blackouts $1-\eta_+$ and $1-\eta_-$ and to recover the true transition rates of this link. A subsequent post-modification of the trajectories yields thermodynamically consistent flow-like observables and combined with an analytical analysis of the TUR and the multi-dimensional TUR (MTUR) lets us recover the full value of the TUR-ratio, and thus the lower bound on entropy production that we would have found in the absence of any blackouts. Finally, the waiting-time distributions based on the post-processed data yield a lower bound on entropy production that can even be stronger than the full TUR.

\begin{figure*}
    \centering
    \includegraphics[scale=1]{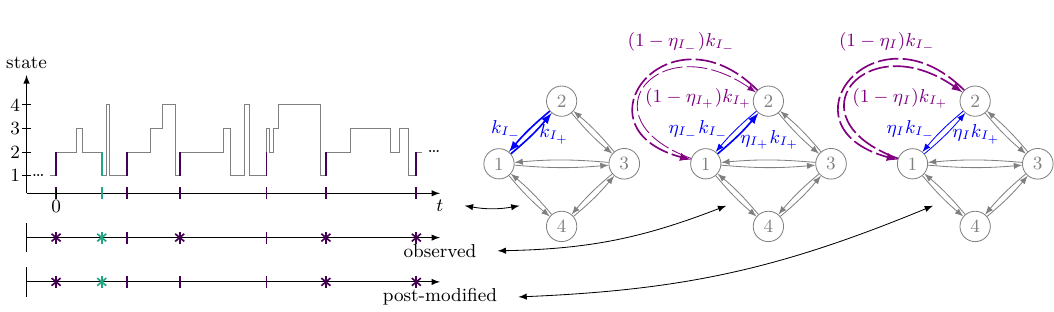}
    \caption{Scheme of observation and post-modification. In the fluctuating microscopic trajectory (gray) of the left 4-state Markov network only transitions $I_+ =(12)$ (indigo) and $I_-=(21)$ (green) with rates $k_{I_+}$ and $k_{I_-}$, respectively, are, in principle, visible to an observer as indicated by the vertical bars. The actually observed transitions indicated by the crosses miss some of these bars due to random blackouts. The detection probabilities $\eta_{I_+}<1$ and $\eta_{I_-}<1$ lead to decreased effective rates $\eta_{I_+}k_{I_+}$ and $\eta_{I_-}k_{I_-}$ and to a second channel (purple) representing the blackouts with rates $(1-\eta_{I_+})k_{I_+}$ and $(1-\eta_{I_-})k_{I_-}$ as indicated in the middle network. The post-modification randomly discards transitions from the observed trajectory such that it corresponds to a virtual dynamics with equal detection probability for forward and backward transitions given by $\eta_I\leq\etastar{I}=\min\{\eta_{I_+},\eta_{I_-}\}$ as indicated by the right network. \re{Thus, observed transitions $I_+$ and $I_-$ are discarded with probability $\eta_{I}/\eta_{I_+}$ and $\eta_{I}/\eta_{I_-}$, respectively.}}
    \label{fig:ps_traj_graphs}
\end{figure*}
\paragraphLetter{Setup}
We assume to observe one or more pairs of a transition and its reverse, which we call links, of a discrete-state Markov network. In particular, if a transition $I_+$ of some link $I=(I_+,I_-)$ occurs, we detect it with probability $\eta_{I_+}$. The same accounts for the corresponding transition $I_-$ with $\eta_{I_-}$. This inability to detect all jumps along a trajectory renders observations incomplete as in the paradigmatic example shown in \Cref{fig:ps_traj_graphs}. The resulting observable dynamics is a renewal process for which the underlying Markov network effectively has two channels for each incompletely observable transition, e.g. a channel for transition $I_+=(ij)$ from state $i$ to $j$ with rate $k_{I_+}\eta_{I_+}$ and one with $(1-\eta_{I_+})k_{I_+}$ as shown in \Cref{fig:ps_traj_graphs}. It is governed by observed, i.e. experimental, waiting-time distributions $\wtdpsiexp{I_r}{J_s}{t}$. These provide the probability density for an observed trajectory that starts after transition $I_r$, with $r=+$ or $-$, of the link $I=(I_+,I_-)$ at time 0 and ends with the next observed transition $J_s$, with $s=+$ or $-$, of the link $J=(J_+,J_-)$ at time $t$. Whereas the distributions $\wtdpsiexp{I_r}{J_s}{t}$ implicitly depend on the full set of detection probabilities $\alletas = \{\eta_{I_+},\eta_{I_-},\eta_{J_+},\eta_{J_-},\dots\}$, the $\wtdpsitheo{I_r}{J_s}{t}$ for the theoretical case of perfect detection do not. Yet, the mean frequency $\nrateexp{I_r}$ of transitions $I_r$ derived from the former differs only by a factor of $\eta_{I_r}$ from its value in the latter case. Furthermore, for each type of superscript $\text{ss}\in\{\text{ex},\text{th}, \superscriptProcessing\text{ (as introduced below)}\}$, waiting-time distributions $\wtdpsiss{I_r}{J_s}{t}$ are normalized such that $\sum_{J_s}\int_0^\infty\wtdpsiss{I_r}{J_s}{t}=1$, where $J_s$ runs over all observable transitions including $I_r$.

Underlying the observable dynamics, we assume a discrete state Markov network in its steady state. Transition rates $k_{ij}>0$ for a jump from state $i$ to $j$ imply $k_{ji}>0$ and generate the dynamics via the master equation
\begin{align}
\partial_t p_i(t) = \sum_{j} [p_j(t)k_{ji} - p_i(t)k_{ij}] \equiv -\sum_{j} j_{ij}(t)
\label{eq:masterEq}
\end{align}
for the probability $p_i(t)$ of the system to be in state $i$ at time $t$. If the system is in equilibrium, \re{i.e., in detailed balance,} all mean net currents $j_{ij}^s$ vanish. Otherwise, the system is in a nonequilibrium steady state (NESS) with steady-state distribution $p_i^s$.

\paragraphLetter{Inference of blackouts and restoring transition rates}
The waiting-time distributions for consecutively observed transitions contain the topological information of how many hidden transitions are in the shortest path between the two observable ones \cite{vdm22,maie24}. \re{In the short-time limit, WTDs are given by the probability density of such shortest path(s), i.e., a factor of $t$ for each hidden transition multiplied with the product of rates along the respective path.} For an incompletely observed link $I$ with $\eta_{I_+}$ and $\eta_{I_-}$, these paths contain no hidden transition between $I_+=(ij)$ and $I_-=(ji)$ and one hidden transition in the sequences $I_+\to I_+$ and $I_-\to I_-$, which\re{, hence, leads to}
\begin{align}
    \wtdpsiexp{I_+}{I_+}{t} &= (1-\eta_{I_-})k_{I_-}\eta_{I_+}k_{I_+}t + O(t^2), \label{eq:stl_pp} \\
    \wtdpsiexp{I_+}{I_-}{t} &= \eta_{I_-}k_{I_-} + O(t), \label{eq:stl_pm} \\
    \wtdpsiexp{I_-}{I_+}{t} &= \eta_{I_+}k_{I_+} + O(t), \label{eq:stl_mp} \\
    \wtdpsiexp{I_-}{I_-}{t} &= (1-\eta_{I_+})k_{I_+}\eta_{I_-}k_{I_-}t + O(t^2). \label{eq:stl_mm}
\end{align}
In the short-time limit, the four equations \eqref{eq:stl_pp} to \eqref{eq:stl_mm} can be solved for the detection probability
\begin{align}
    \eta_{I_+} &= \lim_{t\to 0}\frac{\wtdpsiexp{I_-}{I_+}{t}}{\wtdpsiexp{I_-}{I_+}{t} + \wtdpsiexp{I_-}{I_-}{t}/t\wtdpsiexp{I_+}{I_-}{t}}
\end{align}
and similarly for $\eta_{I_-}$ by exchanging $+$ and $-$, and hence for the probabilities $1-\eta_{I_\pm}$ of a blackout and thereafter for the full rates $k_{I_+}=k_{ij}$ and $k_{I_-}=k_{ji}$. If all links of the network are observable and subject to random blackouts, this scheme yields all transition rates, i.e., the full generator, of the Markov network and hence access to the full dynamics. Using the steady-state probabilities, e.g., obtained as the eigenvector of the generator or by $p_i^s = \nrateexp{I_+}/\eta_{I_+}k_{I_+}$, enables us to check whether two states that are connected by an observable pair of transitions are in detailed balance, i.e., whether $j_{ij}^s=0$. Hence, we can discern equilibrium from a NESS even if all transitions are subject to random blackouts of arbitrary unknown frequencies.

Even if\re{, in contrast to all links as described in the last paragraph,} only a few pairs of transitions are incompletely observable and others are hidden, this scheme allows us to infer the full set $\alletas$ and the rates of detectable transitions \re{$I_r$} as well as, based on their mean rates \re{$\nrateexp{I_r}$}, the steady-state probabilities of their adjacent states. Again, we can discern driven transitions and the ones with zero-mean net current and, if at least one link in each fundamental cycle is observable, effectively driven and undriven cycles.

\begin{figure}
    \centering
    \includegraphics[scale=1]{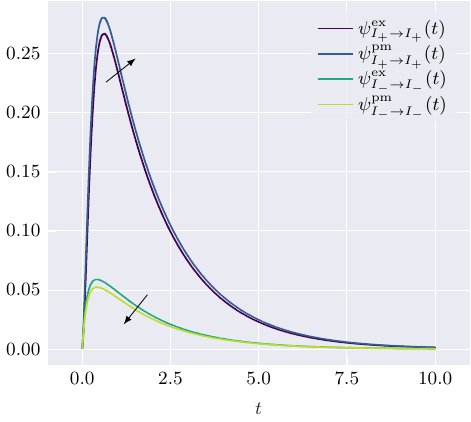}
    \caption{Waiting-time distributions for one observable pair of transitions $I_+=(12)$ and $I_-=(21)$ in the 4-state Markov network from \Cref{fig:ps_traj_graphs} with $k_{12} =  k_{23} = 3$, $k_{31} = k_{14} = k_{43} = 2$ and $k_{13} = k_{21} = k_{32} = k_{34} = k_{41} = 1$ obtained via an analog of Gillespie's algorithm \cite{gill77} for semi-Markov processes\re{, see \hyperref[sec:appMethods]{Methods in End Matter} for details,} with detection probabilities $\eta_{I_+}=0.8$ and $\eta_{I_-}=0.9$, trajectory length $T=7\re{\times} 10^9$, bin width $\Delta t=10^{-3}$ and cut-off time $t_\text{c}=20$. The post-modification illustrated in \Cref{fig:ps_traj_graphs} leads to the distributions for $\eta_I=0.8$ as indicated by the arrows.}
    \label{fig:ps_wtd}
\end{figure}

\paragraphLetter{Post-modifying trajectories}
Knowing the full transition rates $k_{I_\pm}$ and the probabilities $\eta_{I_\pm}$ is insufficient for any inference technique based on higher moments than the mean of some flow, like for the TUR and for the entropy estimator based on waiting-time distributions, since these moments cannot be restored in general. Hence, we need to modify the statistics to achieve consistency between thermodynamic characteristics of the system and observables by removing the bias $\eta_{I_+}\neq\eta_{I_-}$ in all links $I$. Additionally, we use that thermodynamic inference allows for hidden links that we interpret here as the hidden second channel resulting from blackouts of an observable link, see \Cref{fig:ps_traj_graphs}. The goal of \re{post-modifying} the observed trajectory is to generate \re{a trajectory of} a virtual dynamics that is thermodynamically consistent \re{and} for which extant (thermodynamic) inference methods like the TUR estimator can be used. Indeed, if the relation $\eta_{I_+}=\eta_{I_-}$ held true for all $I$, then the blackouts could be considered as a thermodynamically consistent second unobserved channel between the corresponding states. The observations obtained for this particular set $\alletas$ could be used for inference of, e.g., a lower bound on the entropy production of the underlying Markov network by using the TUR. For the generic case of $\eta_{I_+}\neq\eta_{I_-}$, we identify $\etastar{I} \equiv \min\{\eta_{I_+},\eta_{I_-}\}$. We can now randomly discard transitions in the observed trajectory such that the modified trajectory effectively corresponds to $\eta_{I_+}=\eta_{I_-}=\eta_I=\etastar{I}$ as illustrated in \Cref{fig:ps_traj_graphs}. In fact, we can thus generate trajectories for any $\eta_I\leq\etastar{I}$. The example in \Cref{fig:ps_wtd} shows the effect of post-modification on waiting-time distributions. These trajectories can, e.g., be used as input for both the TUR and the entropy estimator based on waiting-time distributions.

\paragraphLetter{Entropy estimators}
\begin{figure*}
    \centering
    \includegraphics[scale=1]{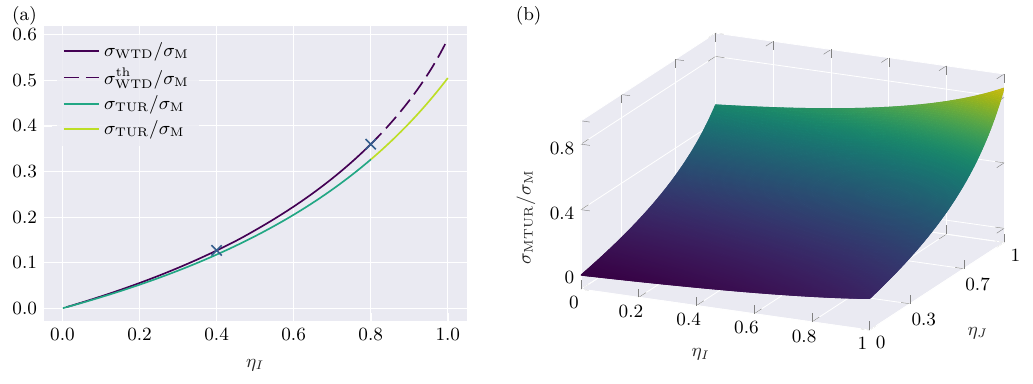}
    \caption{Entropy estimators for observations with blackouts of the 4-state Markov network from \Cref{fig:ps_traj_graphs}. (a) Entropy estimators $\eprTUR(\eta_I)$ and $\eprSM(\eta_I)$ for the rates as given in the caption of \Cref{fig:ps_wtd} as a function of the detection probabilities $\eta_{I_+}=\eta_{I_-}=\eta_I$ in units of the mean entropy production rate $\eprM$ of the underlying Markov network. The waiting-time distributions obtained from a simulated trajectory as presented in \Cref{fig:ps_wtd} result in the highlighted values of $\eprSM(\eta_I)$ (blue crosses). \re{The uncertainty of these two values of $\eta_I$ are not explicitly shown as the corresponding inferred values deviate by a few per mille from their theoretical value, e.g., $<0.38\%$ for the right cross.} The theoretical values of $\eprSM(\eta_I)$ for $\eta_I>\etastar{I}=0.8$ (dashed) are inaccessible and require full knowledge of the system, whereas $\eprTUR(\eta_I)$ can be inferred in this regime through \Cref{res:TUR} (greenyellow). (b) Entropy estimator $\eprMTUR(\eta_I,\eta_J)$ as a function of the detection probabilities $\eta_I$ and $\eta_J$ in units of $\eprM$ of this Markov network with rates $k_{12} =  k_{23} = k_{31} = 2$, $k_{13} = k_{21} = k_{32} = k_{34} = k_{41} = 1$ and $k_{14} = k_{43} = 3$ in which, besides $I_+=(12)$ and $I_-=(21)$, transitions $J_+=(34)$ and $J_-=(43)$ are observable. \re{See \hyperref[sec:appMethods]{Methods in End Matter} for more details on how these data have been generated.}}
    \label{fig:eprEstimators_TUR_MTUR}
\end{figure*}
The mean entropy production rate \cite{peli21,shir23,seif25}
\begin{align}
    \eprM = \sum_{ij}p_i^sk_{ij}\ln\frac{k_{ij}}{k_{ji}}
    \label{eq:eprMarkov}
\end{align}
is not directly accessible in Markov networks with hidden transitions and incomplete detections. However, estimators that remain valid for Markov networks with a hidden channel parallel to each observable link\re{,} like the TUR \re{\cite{liu19},} yield non-trivial lower bounds of $\eprM$. The TUR, $\eprM \geq j^2/D \equiv\eprTUR$ \cite{bara15,ging16}, for one link $I$ is based on its mean current $\currentps{I}$ and the corresponding diffusion coefficient $D_I^\superscriptProcessing$. In the presence of blackouts, the TUR estimator has the universal form
\begin{align}
    \eprTUR(\eta_I) = (\currentps{I})^2/D_I^\superscriptProcessing = a\eta_I/(1+b\eta_I)
    \label{res:TUR}
\end{align}
with real constants $a,b$ as argued below. It is thus sufficient to generate data for $\eta_I=\etastar{I}$ and, e.g., for $\eta_I=\etastar{I}/2$ to determine $a$ and $b$. The numerical value $\eprTUR(1) = a/(1+b)$ then yields the value one would have determined if there were no blackouts.

The universal form of the TUR estimator \eqref{res:TUR} stems from the linear scaling of $\currentps{I}$ with $\eta_I$ and from the intrinsic dynamical fluctuations and the binomial detection noise that lead to two effective terms of the diffusion coefficient $D_I^\superscriptProcessing$, one is linear and one is quadratic in $\eta_I$. Theoretically, we find these dependencies by using the first cumulants extracted from the dominant eigenvalue $\lambda(z)$ of the tilted generator that generates the dynamics of the cumulant generating function \cite{koza99}\re{, as detailed in \hyperref[sec:appendix]{End Matter}}.

For a general current based on multiple observable \re{links $I,J,\dots$}, the TUR estimator takes the form
\begin{align}
    \eprTUR(\alletas) = \re{\frac{(\sum_I a_I\eta_I)^2}{\sum_I (b_I\eta_I + c_I\eta_I^2 ) + \sum_{I<J} d_{IJ}\eta_I\eta_J}},
    \label{res:TUR_multiple}
\end{align}
where \re{$a_I,b_I,c_I,d_{IJ}$} are real \re{and $I<J$ means that each unordered pair $IJ$ contributes one term}. In both cases, the post-modification allows us to determine the estimator $\eprTUR$ for \re{all $\alletas$}, where $0\leq \eta_I \leq\etastar{I}$ for each pair of transitions $I_+,I_-$. After determining its constants\re{, i.e., the $a_{(I)},b_{(I)}(,c_{I}$ and $d_{IJ})$, by fitting its functional form \eqref{res:TUR} or \eqref{res:TUR_multiple}, or the $\eta$-dependence of the contributing cumulants, to the data, as described in \hyperref[sec:appFitting]{End Matter}}, we \re{can} obtain the value of $\eprTUR$ for \re{any $\alletas$} including the ones with $\etastar{I} \leq\eta_I \leq 1$ for all $I$. Notably, we are therefore able to get $\eprTUR$ as if we were able to register all transitions along observable transitions without blackouts, which we illustrate in \Cref{fig:eprEstimators_TUR_MTUR}\,(a).

The MTUR, which yields a tighter bound on $\eprM$ than the TUR for a linear combination of observed currents by accounting for covariances \cite{dech18}, is still valid in our setup and can be analyzed analogously. For two pairs of observable transitions \rev{$I_+,I_-$} and \rev{$J_+,J_-$} with $\eta_I$ and $\eta_J$, the explicit expression for $\eprMTUR$ as a function of $\boldsymbol{\eta}$ becomes
\begin{align}
    \eprMTUR(\eta_I,\eta_J) = \frac{a_1\eta_I + a_2\eta_J + a_3\eta_I\eta_J}{1 + a_4\eta_I + a_5\eta_J + a_6\eta_I\eta_J}
    \label{res:MTUR_twoLinks}
\end{align}
with real constants $a_l,\ l\in\{1,2,...,6\}$ \re{as detailed in \hyperref[sec:appendix]{End Matter}}. As for the TUR, after obtaining these constants from a fit, we can extend this estimator to get its value for $\alletas=\{1,1,\dots\}$ as if we were able to detect all jumps along observable transitions. We illustrate the estimator \eqref{res:MTUR_twoLinks} for a 4-state network in \Cref{fig:eprEstimators_TUR_MTUR}\,(b).

Finally, in addition to the TURs, using waiting-time distributions obtained from our post-modification yields the entropy estimator
\begin{align}
    \eprSM(\alletas) = \sum_{r,s=\pm}\sum_{\rev{I,J}}\int_0^\infty \nrateps{I_r}\wtdpsips{I_r}{J_s}{t}\ln\frac{\wtdpsips{I_r}{J_s}{t}}{\wtdpsips{\widetilde{J_s}}{\widetilde{I_r}}{t}}\dd{t}
    \label{res:eprEstimator_WTD}
\end{align}
that follows from the one derived in Refs. \cite{vdm22,haru22} for complete blackout-free observations of Markov networks with only one pair of transitions between two adjacent states and with the tilde indicating the reversed transition. The estimator \eqref{res:eprEstimator_WTD} constitutes a non-trivial lower bound of $\eprM$ that depends on the full set $\alletas$ in a complex manner that includes infinite sums as the $\wtdpsips{I_r}{J_s}{t}$ do. For an example, we show $\eprSM$, which we have determined analytically and from a numerically simulated trajectory, in \Cref{fig:eprEstimators_TUR_MTUR}\,(a).

\paragraphLetter{Concluding perspective}
We have introduced a strategy to eliminate effects of random blackouts in the detection of transitions in Markov networks. By analyzing the short-time limit of the four waiting-time distributions associated with each link, both the unknown blackout rates and the true transition rates of this link can be inferred. In particular, a restored non-zero current through this link signifies non-equilibrium. If all transitions of a network are observable subject to blackouts with arbitrary unknown frequencies, the full dynamics can thus be restored. By post-modification of the trajectory data in partially observed networks, we can generate data that correspond to time-symmetric blackout rates and that are thermodynamically consistent with the underlying Markov network. Combined with the analytically derived dependence of the TUR-bound on these blackout rates, this approach allows determination of the full bound even in the presence of arbitrary unknown blackouts. The same procedure works for the MTUR. Moreover, we have shown that a lower bound on entropy production derived from waiting-time distributions remains valid when using post-modified trajectories, despite time-asymmetric errors in the original data. \re{The main results derived in this Letter have the status of proven theorems based, in principle, on infinite statistics. It will be important to address the impact of finite statistics \rev{including a quantification of uncertainties, arising, e.g., in the case of undersampling,} in future works ideally in relation with specific experimental data.}

The two main strategies introduced in this Letter -- the recovery of blackout and transition rates and the post-modification of trajectory data based on the inferred blackout rates -- are expected to be applicable beyond the stationary-state scenario discussed here. In particular, the same methodology could be extended to analyze blackouts in periodically and arbitrary time-dependently driven systems. Likewise, this approach of post-processing trajectories should work in the inference of further thermodynamic quantities like cycle affinities \cite{vdm22,ohga23}\re{, of topological information \cite{maie25,zhao25}} and also beyond inequalities based on the TURs and the WTDs and beyond the assumption of time-reversal symmetric detection errors that could be described as spatial errors or faulty coarse-graining \cite{ferr25,vdm25,bao25}. Indeed, from a broader conceptual perspective, it is encouraging for future studies \re{that we have learned here} how to process data in which time-reversal symmetry is not only systematically broken -- as it is in blackout-free observations of a non-equilibrium system -- but additionally compromised in a random and inhomogeneous way.

%
%
%
%
%
%
%
%
%
%
%
%
%
%
%
%
%
%
%
\appendix\crefalias{section}{appendix}\crefalias{subsection}{appendix}
\bibliography{references.bib}
\re{
\onecolumngrid
\section{End Matter}
\twocolumngrid
\paragraphLetter{Tilted-generator approach to cumulants} \label{sec:appendix}
Cumulants of an arbitrary integrated current, e.g., the net count of transitions through a link, are determined by their cumulant generating function or its scaled version. The latter is the largest (dominant) eigenvalue, $\lambda(z)$, of the tilted generator $\vec{L}(z)$ that rules the time evolution of the so-called end-state resolved moment generating function of a trajectory with length $T\to\infty$, see, e.g., \cite{seif25}. The elements of this generator are given by
\begin{align}
\vec{L}_{ij}(z) = k_{ji}e^{zd_{ji}} - \delta_{ij}\sum_l k_{il},
\end{align}
where the non-vanishing $d_{ij} = -d_{ji}$ is the asymmetric increment of transitions from state $i$ to $j$ contributing to the current of interest and $\delta_{ij}$ is the Kronecker delta. In the limit $z\to 0$, the $n$-th derivative of the largest eigenvalue $\lambda(z)$ of this operator then leads to the $n$-th cumulant of the fluctuating current $\hat{j}$,
\begin{align}
\cumu_n(\hat{j}) = T^{1-n} \partial_z\lambda(z)_{\vert z=0}.
\end{align}
A more direct route than finding $\lambda(z)$ and taking its derivatives uses $\lambda(0)=0$ in derivatives of the characteristic polynomial
\begin{align}
\chi(z) = \det{\vec{L}(z) - \lambda\vec{1}} \equiv \sum_{i=0}^d c_i(z)\lambda^i(z) = 0, \label{appeq:char_polynom}
\end{align}
where the unit matrix $\vec{1}$ and $\vec{L}(z)$ have $d$ rows and the last equality enforces $\lambda(z)$ to be an eigenvalue. This approach is iterative leading to the first two cumulants
\begin{align}
j &\equiv \langle\hat{j}\rangle = - \left.c'_0/c_1\right._{\vert z=0} \qq{and} \label{appeq:mean_c} \\
D_{\hat{j}} &= \Var(\hat{j})T/2 = \left.-(c''_0 + 2c'_1j + 2c_2 j^2)/2c_1\right._{\vert z=0}, \label{appeq:var_c}
\end{align}
where primes indicate derivatives $\partial_z$ and we have dropped the $z$-dependencies.

Blackouts modify the elements of the tilted generator to
\begin{align}
\vec{L}_{ij}(z) = k_{ji}(\eta_{ji}e^{zd_{ji}} +1-\eta_{ji}) - \delta_{ij}\sum_l k_{il}. \label{appeq:tiltedGenEta}
\end{align}
For the virtual dynamics after post-modification, we have $\eta_{ij} = 0$ for hidden transitions and $\eta_{ij} = \eta_{ji}$ otherwise. The functions $c_i(z)$ in the characteristic polynomial \eqref{appeq:char_polynom} then depend on the full vector of detection probabilities $\vec{\eta}$. In particular, they consist of terms with factors of the form $k_{ji}(\eta_{ji}e^{zd_{ji}} +1-\eta_{ji})$, which reduce to $k_{ji}$ for $z= 0$. Hence, all first derivatives $c'_i(0)$ either vanish or are linear in the elements of $\vec{\eta}$. Moreover, second derivatives $c''_i(0)$ either vanish or consist of linear terms in at least some $\eta_{ij}$ as well as of terms that are quadratic or bilinear in elements of $\vec{\eta}$. This pattern continues for higher derivatives $c^{(n)}_i(0)$ as a direct consequence of the derivative of products until $n$ equals the largest degree of $c_i(z)$ in $\alletas$. Derivatives $c^{(m)}_i(0)$ with $m>n$ meet the same criteria as the $c^{(n)}_i(0)$. Using these insights, we find that the mean current \eqref{appeq:mean_c} is linear in all elements of $\vec{\eta}$ as one would intuitively expect, i.e.,
\begin{align}
j^\superscriptProcessing = \sum_I a_I \eta_I \qq{with real constants $a_I$.} \label{appeq:meancurrent_eta}
\end{align}
The diffusion coefficient \eqref{appeq:var_c} then becomes
\begin{align}
D_{\hat{j}}^\superscriptProcessing = \sum_I \left(b_I\eta_I + c_I\eta^2_I\right) + \sum_{I<J}d_{IJ}\eta_I\eta_J, \label{appeq:diffCoeff_eta}
\end{align}
where $b_I,c_I$ and $d_{IJ}$ are real constants. We have simplified the notation by using the concept of links $I=(I_+,I_-)$ introduced in the main text because any pair $ij=I_+$ and $ji=I_-$ has $\eta_{I_+} = \eta_{I_-} = \eta_I$. Based on the insights about $c^{(n)}_i(0)$ and the determination of higher derivatives of $\lambda(z)$ making use of lower-order cumulants or, equivalently, of the lower-order derivatives of $\lambda(z)$ like the ones in Eqs. \eqref{appeq:mean_c} and \eqref{appeq:var_c}, the $n$-th cumulant must be a polynomial in the detection probabilities of degree $n$ without constant term.

The above analysis for the case with blackouts generalizes readily to the case of multiple fluctuating currents. In this case, we introduce a vector $\vec{z}$ with one element per fluctuating current. The tilted generator \eqref{appeq:tiltedGenEta} and its characteristic polynomial are, hence, modified such that the exponential functions scale each transition rate only by the element of $\vec{z}$ corresponding to the same fluctuating current. Then, cumulants of a single current result as above and mixed derivatives like $\partial_{z_1}\partial_{z_2}\lambda(\vec{z})\vert_{\vec{z}=0}$ yield joint cumulants of two or more fluctuating currents. In the case of two fluctuating currents, the cumulants so found yield the entropy estimator of the MTUR \eqref{res:MTUR_twoLinks}.

\paragraphLetter{Fitting cumulants or entropy estimators to data}\label{sec:appFitting}
The constants $a_{(I)},b_{(I)}(,c_{I}$ and $d_{IJ})$ in \Cref{res:TUR,res:TUR_multiple,res:MTUR_twoLinks} can be obtained in various ways by fitting the functional form of these equations to measurement data. While arbitrarily elaborate fitting algorithms, or in the case of \Cref{res:TUR_multiple} a smart choice of $\alletas$ simplifying quadrics, yield the constants, we present here a simpler route. Using $n$ estimator values of distinct $\alletas$ for the $n$ constants in \Cref{res:TUR,res:MTUR_twoLinks} provides a system of linear equations. Solving it determines the sought constants. While a similar approach can be adapted to \Cref{res:TUR_multiple} too, there is an even simpler procedure, which works for all three estimators \eqref{res:TUR} to \eqref{res:MTUR_twoLinks}. This procedure uses the dependence of cumulants of a fluctuating current on $\alletas$ derived above since the estimators are a function of these cumulants. Determining the constants in the expression of an estimator, hence, reduces to finding their value in the respective cumulants. Importantly, these cumulants are linear in these constants and each set of data yielding one estimator value yields one value for the mean current \eqref{appeq:meancurrent_eta} and for the diffusion coefficient \eqref{appeq:diffCoeff_eta}. Hence, solving the system of linear equations for both the mean current and the diffusion coefficient leads to all needed constants.

\paragraphLetter{Methods for Figs. \ref{fig:ps_wtd} and \ref{fig:eprEstimators_TUR_MTUR}}\label{sec:appMethods}
The WTDs shown in \Cref{fig:ps_wtd} have been determined using an analog of Gillespie's algorithm for semi-Markov processes since the observed dynamics consisting of transitions and waiting times in between can be seen as a semi-Markov process with transitions figuring as states. Like in Gillespie's algorithm the next observable transition and the waiting time since the last observed transition are each fixed by a randomly drawn number, which is uniformly distributed in $[0,1)$, based on analytically determined cumulative WTDs. These cumulative WTDs result by integration from the WTDs, which are obtained by solving the absorbing master equation of the Markov network resulting from rerouting observable transitions (channels) into auxiliary states \cite{seki22,vdm22}. Since sampling infinitely many bins is impossible in finite time, we discard any transition with waiting time longer than the cutoff time $t_c$, i.e., we determine the binned WTDs only in the intervall $[0,t_c]$, where $t_c$ is large enough so that the true WTDs obtained via the absorbing network are negligibly small beyond this time.

Besides data obtained from such a Gillespie-like simulation, \Cref{fig:eprEstimators_TUR_MTUR} contains analytically determined data. First, unless stated otherwise, WTDs have been determined using the analytical method described above. Second, cumulants of one or multiple observed fluctuating currents needed for the TUR or the MTUR, respectively, have been determined analytically using the formalism based on the tilted generator, which we describe above. Third, the entropy production rate \eqref{eq:eprMarkov} of the underlying Markov network has been calculated using the transition rates and the steady-state distribution, determined as the eigenvector associated with eigenvalue zero, of the master equation \eqref{eq:masterEq}.
}

\end{document}